\documentclass[11pt]{article}
\usepackage{times}
\usepackage{amsmath,amsfonts,amssymb,latexsym,epsfig}
\usepackage{color,epsf}
\usepackage{graphicx,stix}
\usepackage{alltt}

\newcommand{\e}{\ensuremath{\mathrm{e}}}
\newcommand{\ad}{\ensuremath{\mathrm{ad}}}

\begin{document}
\title{A general formula for the Magnus expansion in terms of iterated integrals of right-nested commutators}

\author{Ana Arnal\thanks{Email: \texttt{ana.arnal@uji.es}}
 \and
 Fernando Casas\thanks{Corresponding author. Email: \texttt{Fernando.Casas@uji.es} }\and Cristina Chiralt\thanks{Email: \texttt{chiralt@uji.es}}
  }

\maketitle

\begin{abstract}

We present a general expression for any term of the Magnus series as an iterated integral of a linear combination of  independent right-nested commutators with
given coefficients.  The relation with the Malvenuto--Reutenauer Hopf algebra of permutations is also discussed.

\vspace*{1cm}

\begin{center}
Institut de Matem\`atiques i Aplicacions de Castell\'o (IMAC) and  De\-par\-ta\-ment de
Ma\-te\-m\`a\-ti\-ques, Universitat Jaume I,
  E-12071 Cas\-te\-ll\'on, Spain.
\end{center}

\end{abstract}\bigskip



\section{Introduction}

The Magnus expansion allows one to express the fundamental solution of a linear matrix differential equation with varying coefficients as the exponential of an infinite series whose terms
involve time-ordered integrals of nested commutators. In its original formulation \cite{magnus54ote}, it was established as follows. 
Let  $A(t)$ be a known function of $t$ in the ring of all power series of the type 
\[
  A(t) = \sum_{n=0}^{\infty} u_n t^n
\]
and let $Y(t)$ be an unknown function satisfying the initial value problem 
\begin{equation}  \label{eqdif}
Y^{\prime}(t)=A(t)Y(t),\qquad\qquad Y(0)=I.
\end{equation}
Then 
\begin{equation}   \label{me.1}
   Y(t) = \exp \Omega(t),
\end{equation}
where $\Omega$ is an infinite series    
   \begin{equation}
\Omega(t)=\sum_{k=1}^{\infty}\Omega_{k}(t), \qquad \mbox{ with } \qquad \Omega_k(0) = 0  \label{ME}%
\end{equation}
that is obtained by inserting (\ref{me.1}) into  (\ref{eqdif}) and solving the
differential equation satisfied by  $\Omega$:
\begin{equation}   \label{OmegaDE}
  \frac{d \Omega}{dt} = \sum_{n=0}^\infty \frac{B_n}{n!} \,
  {\ad}^n_\Omega A, \qquad \Omega(0) = 0.
\end{equation}
Here
\[
 \mathrm{ad}_{\Omega}^0 A = A, \qquad \mathrm{ad}_{\Omega}^{k+1} A  =
    [ \Omega, \mathrm{ad}_{\Omega}^k A ], \qquad k \ge 0,
\]    
$B_j$ are Bernoulli numbers and $[A,B] \equiv A B - B A$ denotes the usual commutator.

Applying Picard's iteration to (\ref{OmegaDE}) one obtains
\[
  \Omega^{[0]} = 0, \qquad \Omega^{[1]} = \int_0^t A(t_1) dt_1,
\]
\[
  \Omega^{[n]} = \int_0^t \left( A(t_1) dt_1 - \frac{1}{2}
  [\Omega^{[n-1]}, A] + \frac{1}{12}
  [\Omega^{[n-1]},[\Omega^{[n-1]},A]] + \cdots \right)
     dt_1.
\]
leading to the first terms in the series (\ref{ME}) as 
\begin{align}
\Omega_{1}(t)  & =\int_{0}^{t}A(t_{1}) dt_{1},\nonumber\\
\Omega_{2}(t)  & =  -\frac{1}{2} \int_0^t \left[ \int_0^{t_1} A(t_2) dt_2, A(t_1) \right] dt_1 \nonumber \\
\Omega_3(t) & =  \frac{1}{12} \int_0^t \left[ \int_0^{t_1} A(t_2) dt_2, \left[ \int_0^{t_1} A(t_2) dt_2, A(t_1) \right] \right] dt_1 \nonumber \\
  & + \frac{1}{4} \int_0^t \left[ \int_0^{t_1} \left[ \int_0^{t_2} A(t_3) dt_3, A(t_2) \right] d t_2, A(t_1) \right] dt_1  \label{ME123} 
\end{align}
and a more involved expression for $\Omega_4$ (see e.g. \cite{iserles00lgm}). Picard's  theorem then insures 
that $\lim_{n \rightarrow \infty} \Omega^{[n]}(t) = \Omega(t)$ in a neighborhood of $t=0$. By doing some algebra it is possible to write 
down explicitly at least the first $\Omega_k$ as linear combinations of iterated integrals of
nested commutators of $A$ evaluated at different times, but the complexity of this task increases steadily with $k$. For instance,
working out the successive integrals appearing in $\Omega_3$ as given by (\ref{ME123}) we get
\begin{equation}  \label{omh.3}
   \Omega_3(t) =\frac{1}{6}\int_{0}^{t}dt_{1}\int_{0}^{t_{1}}%
dt_{2}\int_{0}^{t_{2}} dt_{3}\ \Big( \left[  A(t_{1}),\left[
A(t_{2}),A(t_{3})\right]  \right]  
 +    \left[  A(t_{3}),\left[  A(t_{2}%
),A(t_{1})\right]  \right]  \Big), 
\end{equation}
whereas similar expressions for $\Omega_4$ and $\Omega_5$ have been presented in \cite{prato97ano}. At any rate, this structure
 is especially favorable in practice when the differential equation evolves in a Lie group and
the series is truncated: the approximation thus obtained still belongs to the same Lie group and thus shares with the exact solution relevant qualitative properties
\cite{iserles00lgm,blanes09tme}.

Since the 1960s the Magnus expansion (often with different names)
has been used to render analytical approximations in many different areas of science, ranging from nuclear, atomic and molecular physics 
to nuclear magnetic
resonance, quantum electrodynamics, control theory, and also as a numerical integrator for differential equations in the realm of geometric numerical integration
(see \cite{blanes09tme} for a review). Here the aim
is to construct integration schemes that preserve the main qualitative features of the exact solution, such as its invariant quantities or the geometric structure 
\cite{blanes16aci,hairer06gni}. The convergence
of the expansion is also an important feature and several results are available in the literature \cite{blanes98maf,casas07scf,moan08cot,lakos17cef}.

Different procedures have been proposed along the years to obtain explicit expressions of $\Omega_k$ for any $k$ in terms of commutators: recurrence relations \cite{klarsfeld89rgo},
techniques based on binary trees \cite{iserles99ots}, combinatorial techniques applied to iterated integrals \cite{mielnik70cat,vinokurov92lot,agrachev94tsp}, etc. 
The expressions thus obtained for $\Omega_k$
present however some limitations: they are not unique (due to the Jacobi identity and other identities appearing at higher orders) and very often not all the terms
are independent. For certain applications it might be of some interest to get expressions similar to (\ref{ME123}) for \emph{any} given $\Omega_k$, i.e.,
writing an arbitrary $\Omega_k$ as an \emph{iterated integral} of (a linear combination of) \emph{independent} nested commutators. As far as we know, this has been carried out only
up to $k=6$ \cite{lakos17cef} and it is the purpose of this paper to provide a general expression for any $k \ge 1$, namely we will provide an explicit formula for $\Omega_k$ as an iterated integral
of a linear combination of $(k-1)!$ right-nested independent commutators of $A$ evaluated at different times. 
In doing so we will relate the Magnus expansion with the well known Malvenuto--Reutenauer Hopf algebra of permutations \cite{hazewinkel10ara}, thus providing a new illustration of this abstract
algebraic structure.

\section{The Magnus expansion in terms of iterated integrals}
\label{sec.2}

As in \cite{strichartz87tcb,agrachev94tsp} our starting point is to write the  Magnus series (\ref{ME}) in terms of iterated integrals of $A$. This can be achieved by considering the 
Neumann series for the solution of (\ref{eqdif}),
\[
   Y(t) = I + \int_0^t A(s) ds +  \int_0^t dt_1 \int_0^{t_1} dt_2 \, A(t_1) A(t_2) + \cdots
\]
or, in general, 
 \begin{equation}   \label{N.1}
    Y(t) = I + \sum_{n=1}^{\infty} P_n(t)
 \end{equation}
 with
 \begin{equation}   \label{Pn}
   P_n(t) = \int_0^t dt_1 \int_0^{t_1} dt_2 \cdots \int_0^{t_{n-1}} dt_n  \, A(t_1) A(t_2) \cdots A(t_n).
 \end{equation}
This is a convergent series for all $t$ (if $A$ is bounded), but, in contrast to the Magnus expansion, when truncated  no longer preserves qualitative properties of the exact
solution. In particular, if $A(t)$ is a skew-Hermitian operator, the approximation thus obtained is no longer unitary.

The $\Omega_k$ can in fact be expressed in terms of the iterated integrals (\ref{Pn}) by taking logarithms in (\ref{ME}) and equating with (\ref{N.1}),
\[
\sum_{k=1}^{\infty }\Omega _{k}(t)=\log \left( I+\sum_{k=1}^{\infty
}P_{k}(t)\right).
\]
Then
\begin{equation}
\Omega
_{n}=P_{n}-\sum_{j=2}^{n}\frac{(-1)^{j}}{j}R_{n}^{(j)},\qquad
n\geq 2, \label{MagDay}
\end{equation}
where
\[
  R_n^{(j)}=\sum P_{i_1}P_{i_2} \cdots P_{i_j}  \qquad
(i_1+\cdots+i_j=n)
\]
and the sum extends over all $i_1, i_2, \ldots i_j$ such that $i_1+i_2 + \cdots+i_j=n$ \cite{burum81meg}.
Thus, in particular, we get for the first terms
\begin{eqnarray}\label{Salzeqs}
  \Omega_1&=& P_1 \nonumber \\
  \Omega_2&=& P_2-\frac{1}{2}P_1^2 \\
  \Omega_3&=& P_3-\frac{1}{2}(P_1P_2+P_2P_1)+\frac{1}{3}P_1^3   \nonumber \\
  \Omega_4 & = & P_4 - \frac{1}{2} (P_1 P_3 + P_3 P_1 + P_2^2) + \frac{1}{3} (P_1^2 P_2 + P_1 P_2 P_1 + P_2 P_1^2) - \frac{1}{4} P_1^4. \nonumber
\end{eqnarray}
Notice, however, that from $\Omega_2$ on, these expressions are not yet written in terms of time-ordered integrals. To achieve this goal we have to express the products
$P_{i_1} \cdots P_{i_j}$ appearing in (\ref{MagDay}) as iterated integrals. In this respect, it is useful to introduce the following notation:
\begin{equation}   \label{permu.it}
  A(i_1 i_2 \ldots i_n ) \equiv \int_0^t dt_1 \int_0^{t_1} dt_2 \cdots \int_0^{t_{n-1}} dt_n \, A(t_{i_1}) A(t_{i_2}) \cdots A(t_{t_n}).
\end{equation}
Observe that the order in the integration is fixed, whereas the ordering of the product appearing in the integrand is indicated by the sequence
$i_1, i_2, \ldots i_n$. According with this notation, we have
\[
\begin{aligned}
 &  A( 1 ) = \int_0^t dt_1 A(t_1), \qquad A( 1 2 ) = \int_0^t dt_1 \int_0^{t_1} dt_2 \, A(t_1) A(t_2) \\
 & A( 2 3 4 1) = \int_0^t dt_1 \int_0^{t_1} dt_2 \int_0^{t_2} dt_3 \int_0^{t_3} dt_4 \, A(t_2) A(t_3) A(t_4) A(t_1)
\end{aligned} 
\]
etc, whereas (\ref{Pn}) simply reads
\[
   P_n(t) = A(1 2 \ldots n). 
\]
Having established a one-to-one correspondence between iterated integrals and permutations via equation (\ref{permu.it}), it is possible to encode the products
appearing in (\ref{MagDay}) also in terms of permutations. Thus, in particular,
\[
\begin{aligned}
  & \Omega_2 = A( 1 2 ) - \frac{1}{2}  A( 1 ) \cdot  A( 1 ) \\
  & \Omega_3 = A(1 2 3) - \frac{1}{2} A(1) \cdot A(1 2) -  \frac{1}{2} A(1 2) \cdot A(1) + \frac{1}{3} A( 1 )  \cdot A( 1 ) \cdot A(1).
\end{aligned}
\]
According with Fubini's theorem, 
\begin{equation}\label{intxy}
   \int_0^\alpha dy \int_y^\alpha f(x,y) \, dx =
  \int_0^\alpha dx \int_0^x f(x,y) \, dy,  
\end{equation}
we have $A( 1 ) \cdot  A( 1 )  =  A( 1 2 )  +  A( 2 1 )$, so that 
\begin{equation}   \label{om.2}
  \Omega_2 =  A( 1 2 ) - \frac{1}{2} \big( A( 1 2 ) + A( 2 1 ) \big) = \frac{1}{2} \big(  A(1 2 ) -  A( 2 1 ) \big). 
\end{equation}
Analogously, using again (\ref{intxy}) one has 
\begin{equation}  \label{prod.1}
\begin{aligned}
  &    A( 1 ) \cdot A(1 2 ) = A(1 2 3 ) + A( 2 1 3 )  + A( 3 1 2 )  \\
  &    A( 1 2 ) \cdot A( 1 )  = A( 1 2 3 )  + A( 1 3 2 )  + A( 2 3 1 ) \\
  & A( 1 ) \cdot A( 1 ) \cdot A( 1 ) =  A( 1 2 3 ) + A( 1 3  2 )  + A( 2 1 3 ) +
     A( 2 3 1 )  + A( 3 1 2 )  + A( 3 2 1 ), 
\end{aligned}
\end{equation}
so that, by inserting (\ref{prod.1}) into the expression of $\Omega_3$ given in (\ref{Salzeqs}), we arrive at
\begin{equation}  \label{om.3}
 \Omega_3 =  \frac{1}{3} \, A(1 2 3) - \frac{1}{6} \, A(1 3 2)  - \frac{1}{6} \, A(2 1 3) - \frac{1}{6} \, A(2 3 1)  - \frac{1}{6}  \, A(3 1 2) + 
 \frac{1}{3} \, A(3 2 1).  
\end{equation}
This procedure can be generalized to higher orders by realizing that any product of integrals encoded in terms of permutations 
appearing in $\Omega_k$
can be replaced by a sum of all possible permutations
of time ordering consistent with whatever time ordering existed within the factors of the original product \cite{dragt83con}. This is again a
consequence of Fubini's theorem (\ref{intxy}). Thus, $A( 1 ) \cdot A( 1 2 )$ in equation (\ref{prod.1}) is the sum of 
all permutations of three elements such 
that the second index is always less than the third
index. On the other hand, in $A( 1 ) \cdot A( 1 ) \cdot A( 1 )$ there is no special ordering, so that there is no preferential order for the
decomposition (and thus all possible permutations have to be taken into account), whereas
\[
  A( 1 ) \cdot A( 1 2 3 )  = A( 4 1 2 3) + A( 3 1 2 4 ) + A( 2 1 3 4 )  + A( 1 2 3 4 ).
\]
By applying this procedure to $\Omega_4$ as given by (\ref{Salzeqs}) we get a linear combination with rational coefficients of all the $4! = 24$ permutations from
the set $\{1,2,3,4\}$. In general, we have for any $n \ge 2$,
\begin{equation}   \label{me.ite1}
  \Omega_n(t) = \sum_{\sigma \in \mathfrak{S}_n} \, (-1)^{d_b} \, \frac{d_a! \, d_b!}{n!} \, A(\sigma),
\end{equation}
where $\sigma \in \mathfrak{S}_n$ denotes a permutation of $\{ 1, 2, \ldots, n \}$, $d_a$ is the number of \emph{ascents} in $\sigma$, $d_b$ is the number of
\emph{descents} and the sum is over the $n!$ permutations of the symmetric group $\mathfrak{S}_n$. We recall that $\sigma$ has an ascent in $i$ if $\sigma(i) < \sigma(i+1)$, $i=1,\ldots, n-1$
and it has a descent in $i$ if $\sigma(i) > \sigma(i+1)$. Here $(i_1 i_2 \ldots i_n) = (\sigma(1) \, \sigma(2) \, \ldots \sigma(n))$. Clearly
$d_a + d_b = n-1$ so that (\ref{me.ite1}) can be written only in terms of either $d_a$ or $d_b$. In this last case one has the alternative expression
\begin{equation}   \label{me.ite2}
  \Omega_n(t) = \frac{1}{n} \sum_{\sigma \in \mathfrak{S}_n} \, (-1)^{d_b} \, \frac{1}{ \binom{n-1}{d_b}} \, A(\sigma),
\end{equation}
or more explicitly
\begin{equation}   \label{me.ite3}
  \Omega(t) = \frac{1}{n} \sum_{n=1}^{\infty} \sum_{\sigma \in \mathfrak{S}_n} \, (-1)^{d_b} \, \frac{1}{ \binom{n-1}{d_b}} \, 
    \int_0^t dt_1 \int_0^{t_1} dt_2 \cdots \int_0^{t_{n-1}} dt_n \, A(t_{\sigma(1)}) A(t_{\sigma(2)}) \cdots A(t_{\sigma(n)}).
\end{equation}
This formula has been published a number of times in the literature, obtained by different techniques \cite{bialynicki69eso,mielnik70cat,strichartz87tcb,agrachev94tsp}.
 If one is interested in writing $\Omega_n$ explicitly as an element
in the Lie algebra generated by the family $A(t)$, the usual approach is then to apply the Dynkin--Specht--Wever theorem \cite{bonfiglioli12tin} to equation 
(\ref{me.ite2}): the resulting expression is obtained by replacing 
\[
      A(t_{\sigma(1)}) A(t_{\sigma(2)}) \cdots A(t_{\sigma(n)}) \quad \mbox{ by } \quad 
 \frac{1}{n} [A(t_{\sigma(1)}), [A(t_{\sigma(2)}), \cdots , [A(t_{\sigma(n-1)}), A(t_{\sigma(n)})] \cdots ]]
\]
in (\ref{me.ite2}). In that case, though, not all the commutators appearing in the corresponding formula are linearly independent
among each other, due to antisimmetry and the Jacobi identity.
By contrast, in the formulation we propose all the terms are
independent.

\section{Iterated integrals and the Hopf algebra of permutations}

The product $A(\sigma) \cdot A(\tau)$, with $\sigma$ and $\tau$ two given permutations, that we introduced in the 
previous section just as a symbolic way of encoding the product of
iterated integrals $P_j$, correspond in fact to a much deeper characterization of the set of permutations. 
This is in fact related with
the Malvenuto--Reutenauer Hopf algebra of permutations, introduced and studied in \cite{malvenuto95dbq,poirier95adh}.
We next briefly recall the construction of this Hopf algebra. In doing so we follow the notation used originally in 
\cite{malvenuto95dbq} for the product(s) and coproduct(s).

By following \cite{aguiar02sot}, let us denote by $\mathfrak{S}Sym$ the graded $\mathbb{Q}$-vector space with fundamental basis 
given by the disjoint union of the symmetric
groups $\mathfrak{S}_n$ for all $n \ge 0$. In particular, $\mathfrak{S}_0 = \{ ( \, ) \}$ and the elements of $\mathfrak{S}_n$ are
considered as words $\alpha = ( a_1 a_2 \ldots a_n )$ on the alphabet $\{1, 2, \ldots, n\}$. In \cite{malvenuto95dbq}
two Hopf algebra structures on $\mathfrak{S}Sym$ are introduced as follows. 

The product $\ast'$ of $\sigma \in \mathfrak{S}_k$ and $\tau \in \mathfrak{S}_{\ell}$ is defined by
\begin{equation}   \label{prod1}
   \sigma \ast' \tau = \sigma \shuffle \bar{\tau},
\end{equation}
where $\bar{\tau}$ is the word in $\{k+1, \ldots, k+ \ell \}$ is obtained by replacing in $\tau$ each $i$ by $i + k$, and
$\shuffle$ denotes the usual shuffle product. Thus, for instance, 
\[
\begin{aligned}
  &   (1)  \ast'  (1 2) =  (1 2 3) +  (2 1 3)  +  (2 3 1) \\
  &   (1) \ast' (2 1) =  (1 3 2)  +  (3 1 2) +  (3 2 1) \\
  &  (1 2) \ast' (1 2) =  (1 2 3 4) +  (1 3 2 4) +  (1 3 4 2) +  (3 1 2 4) +
      (3 1 4 2) +  (3 4 1 2).
\end{aligned}
\]
Notice that the empty word (permutation) acts as the unit element. Given a word $\alpha = (a_1 a_2 \ldots a_m)$ 
without repeats over the alphabet $\{1, 2, \ldots, m\}$, its standardization $\mbox{st}(\alpha)$ is the word 
obtained by applying to $\alpha$ the unique increasing bijection $\{ a_1, \ldots, a_m \} \longrightarrow \{ 1, 2, \ldots, m \}$.
For instance, $\mbox{st}((3 2 4)) = (2 1 3)$ and $\mbox{st}(( \, )) = ( \, )$. Then the coproduct $\delta'$ is defined as
\[
   \delta'(\alpha) = \sum_{\alpha = u v} \mbox{st}(u) \otimes \mbox{st}(v),
\]
where the sum is over all concatenation factorizations of $\alpha$. In particular, 
\[
\begin{aligned}
  & \delta'((2 4 3 1)) = \mbox{st}(( \, )) \otimes \mbox{st}((2 4 3 1)) + \mbox{st}(( 2 )) \otimes \mbox{st}((4 3 1)) + 
       \mbox{st}(( 2 4)) \otimes \mbox{st}((3 1)) \\ 
   & \quad    +  \mbox{st}(( 2 4 3)) \otimes \mbox{st}((1)) + 
        \mbox{st}(( 2 4 3 1)) \otimes \mbox{st}((\,)) \\
  &  =    ( \, ) \otimes (2 4 3 1) + (1) \otimes (3 2 1) + (1 2) \otimes (2 1) + (1 3 2) \otimes (1) + (2 4 3 1) \otimes ( \, ).
\end{aligned}
\]
With the counit defined by $\varepsilon(( \, )) = 1$ and $\varepsilon(\alpha) = 0$ if $\alpha$ has length $\ge 1$, 
$\mathfrak{S}Sym$ is a non-commutative and non-cocommutative Hopf algebra, graded by the length of permutations.

As a matter of fact, another product $\ast$ and another coproduct $\delta$ can be defined endowing $\mathfrak{S}Sym$ with a second
Hopf algebra structure which happens to be isomorphic to the previous one.
Given, as before, $\sigma \in \mathfrak{S}_k$ and $\tau \in \mathfrak{S}_{\ell}$,
\begin{equation}  \label{prod2}
   \sigma \ast \tau = \sum u v,
\end{equation}
where the sum is over all $u$, $v$ such that $\mathrm{st}(u) = \sigma$, 
$\mathrm{st}(v) = \tau$ and the concatenated word $uv$ is a permutation in $\mathfrak{S}_{k + \ell}$.
 Thus, for instance,
\[
\begin{aligned}
  &   (1)  \ast  (1 2) =  (1 2 3) +  (2 1 3)  +  (3 1 2) \\
  &   (1) \ast (2 1) =  (1 3 2)  +  (2 3 1) +  (3 2 1) \\
  &  (1 2) \ast (1 2) =  (1 2 3 4) +  (1 3 2 4) +  (1 4 2 3) +  (2 3 1 4) +
      (2 4 1 3) +  (3 4 1 2).
\end{aligned}
\]
Denoting by $\alpha_{B}$ the word obtained from $\alpha$ by removing all letters that are not in $B$, the coproduct is
defined as
\[
  \delta(\alpha) = \sum_{i=0}^n \alpha_{\{1, \ldots, i \}} \otimes \mbox{st}(\alpha_{\{i+1, \ldots, n\}}).
\]
For example,
\[
\begin{aligned}
  & \delta((2 4 3 1)) = ( \, ) \otimes \mbox{st}((2 4 3 1)) + ( 1 ) \otimes \mbox{st}((2 4 3)) + 
       ( 2 1) \otimes \mbox{st}((4 3)) \\ 
   & \quad    +  ( 2  3 1) \otimes \mbox{st}((4)) + 
        ( 2 4 3 1) \otimes \mbox{st}((\,)) \\
  &  =    ( \, ) \otimes (2 4 3 1) + (1) \otimes (1 3 2) + (2 1) \otimes (2 1) + (2 3 1) \otimes (1) + (2 4 3 1) \otimes ( \, ).
\end{aligned}
\]
These two graded Hopf algebras on $\mathfrak{S}Sym$ are isomprhic and dual to each other (i.e., self-dual)
with respect to the inner product 
$\langle \, , \rangle$ defined by stating that the basis of $\mathfrak{S}Sym$ consisting of all permutations to be orthogonal \cite{poirier95adh,hazewinkel10ara}.
Moreover, if $\theta: \mathfrak{S}Sym \longrightarrow \mathfrak{S}Sym$ denotes the linear involution that takes a
permutation $\sigma$ to its inverse,   $\theta(\sigma) = \sigma^{-1}$, then these two Hopf algebras are conjugated by $\theta$:
\begin{equation}   \label{p1p2}
   \sigma \ast \tau = \theta \big( \theta (\sigma) \ast' \theta (\tau) \big) \quad \mbox{ and } \quad
   \delta (\sigma) = (\theta \otimes \theta) \Big( \delta' \big(\theta(\sigma)\big) \Big).
\end{equation}
We notice at once the connection between the product of iterated integrals arising from the application of Fubini's theorem
(see e.g. (\ref{prod.1})) and the product $\ast$ of
permutations in the Hopf algebra $\mathfrak{S}Sym$  via the one-to-one correspondence between iterated 
integrals and permutations (\ref{permu.it}): it clearly holds that
\begin{equation}   \label{asig}
   A(\sigma) \cdot A(\tau) = A(\sigma \ast \tau).
\end{equation}   
Relation (\ref{asig}) can be found in reference \cite{agrachev94tsp}, where the product $\ast$ is referred to as shuffle product
of permutations.

One might ask what is the equivalent, at the level of iterated integrals, of the product $\ast'$ in $\mathfrak{S}Sym$. To this end, we remark that it is possible to
define another one-to-one correspondence between iterated integrals and permutations, in addition to (\ref{permu.it}). Specifically, let us denote
\begin{equation}   \label{permut.2}
   A'(i_1 i_2 \cdots i_n) \equiv \int_0^t dt_{i_1} \int_0^{t_{i_1}} dt_{i_2} \cdots \int_0^{t_{i_{n-1}}} dt_{i_n} \, A(t_{1}) A(t_{2}) \cdots A(t_n),
\end{equation}   
so that the indices of the permutation indicate the simplex in which the integration is carried out, whereas the order in the functions appearing in the integrand is
fixed. Then, it is straightforward to verify that
\[
   A'(i_1 i_2 \cdots i_n) =  A \big(( i_1 i_2 \cdots i_n )^{-1} \big) = A \big( \theta(i_1 i_2 \cdots i_n) \big).
\]
Thus, the product of iterated integrals of the form  (\ref{permut.2}) corresponds precisely to the product $\ast'$ in $\mathfrak{S}Sym$, and the map $\theta$
relates both types of iterated integrals, i.e.,
\begin{equation}   \label{asig2}
   A'(\sigma) \cdot A'(\tau) = A'(\sigma \ast' \tau).
\end{equation}

\section{Magnus series in terms of right-nested independent commutators}

The algorithm based on the application of (\ref{MagDay}), the product of permutations $\ast$ and the relation (\ref{asig})
 allows us to construct  $\Omega_n$ in the Magnus series explicitly in terms of elements in $\mathfrak{S}Sym$ for any $n \ge 1$. 
We next show that, by appropriately manipulating the expression (\ref{me.ite2}), it is possible to write $\Omega_n$ in such
a way that only right-nested independent commutators are present. 

To illustrate the procedure, consider again the expressions of $\Omega_2$ and $\Omega_3$ given by (\ref{om.2}) and
(\ref{om.3}), respectively. It is clear that (\ref{om.2}) already corresponds to the formula collected in (\ref{ME123}) for $\Omega_2$,
or equivalently
\[
  \Omega_{2}(t)   = - \frac{1}{2}\int_{0}^{t} dt_{1} \int_{0}^{t_{1}}
dt_{2}\ \left[  A(t_{2}),A(t_{1})\right], 
\]
whereas (\ref{om.3}) can be written as
\[
\begin{aligned}
  &  \Omega_3 =   \frac{1}{6} \big( A(1 2 3) - A(2 1 3) \big) - \frac{1}{6} \big( A(2 3 1) - A(3 2 1) \big) - \frac{1}{6} \big( A(3 1 2) - A(3 2 1) \big) \\
   & \quad \; +  \frac{1}{6} \big( A(1 2 3) - A(1 3 2) \big),
\end{aligned}
\]
i.e.,
\[
\begin{aligned}
 &  \Omega_3 = \frac{1}{6}   \int_0^t dt_1 \int_0^{t_1} dt_2 \, \int_0^{t_2} dt_3 [A(t_1), A(t_2)] A(t_3) \\
 & - \frac{1}{6}   \int_0^t dt_1 \int_0^{t_1} dt_2 \, \int_0^{t_2} dt_3  [A(t_2), A(t_3)] A(t_1) \\
 & - \frac{1}{6}   \int_0^t dt_1 \int_0^{t_1} dt_2 \, \int_0^{t_2} dt_3 A(t_3) [A(t_1), A(t_2)] \\
 & + \frac{1}{6}   \int_0^t dt_1 \int_0^{t_1} dt_2 \, \int_0^{t_2} dt_3 A(t_1) [A(t_2), A(t_3)], 
\end{aligned}
\]
whence the expression (\ref{omh.3}) is recovered. Alternatively, Jacobi identity allows us to write also
\[
\begin{aligned}
 &  \Omega_3 = \frac{1}{3}  \int_0^t dt_1 \int_0^{t_1} dt_2 \int_0^{t_2} dt_3 \,  [A(t_3), [A(t_2), A(t_1)]] \\
 &  -  \frac{1}{6}  \int_0^t dt_1 \int_0^{t_1} dt_2 \int_0^{t_2} dt_3 \,  [A(t_2), [A(t_3), A(t_1)]].
\end{aligned} 
\] 
If we denote, in general, 
\[
  A[i_1, i_2, \ldots, i_n] \equiv \int_0^t dt_1 \int_0^{t_1} dt_2 \cdots \int_0^{t_{n-1}} dt_n [A(t_{i_1}), [A(t_{i_2}), \cdots [A(t_{i_{n-1}}), A(t_{i_n})] \cdots ]]
\]
then we can write in a more compact way
\[
 \Omega_2 = -\frac{1}{2} A[2,1], \qquad\qquad  \Omega_3 = \frac{1}{3} \, A[3,2,1] - \frac{1}{6} \, A[2,3,1].
\]  
For higher order terms the same strategy can be applied, namely we can expand (\ref{me.ite2}) and then collect the resulting 
terms into multiple commutators, although the procedure is cumbersome for $n \ge 4$. We rely instead in results
presented in \cite{dragt83con} concerning the set of all  $(N-1)$-fold commutators of
 $N$ different (abstract) linear operators $O_1, O_2, \ldots, O_N$. Specifically, in the Appendix of \cite{dragt83con} it is shown that
\begin{itemize}
  \item[(1)] This set forms a vector space of dimension $(N-1)!$.
  \item[(2)] A possible basis for this vector space is formed by right-nested commutators of the form 
  \[
      [O_m, [O_{\ell},\ldots [O_k, O_j] \ldots ]].
  \]
  \item[(3)] In forming such a basis we can use only those right-nested commutators ending with a 
  particular but otherwise arbitrary operator selected from
  the collection $O_1, O_2, \ldots, O_N$. If we choose this operator as $O_1$, then the basis is formed by the right-nested commutators of the form  
  \[
      [O_k, [O_{j},\ldots [O_i, O_1] \ldots ]],
  \]
  where the indices $k, j, \ldots i$ are all possible permutations of $\{2, 3, \ldots N \}$ (clear\-ly,  $(N-1)!$ 
  permutations).
  \item[(4)] Consider an
  expression which is known to be decomposable into a set of $(N-1)$-fold commutators of $N$ objects and suppose   
  all the right-nested commutators ending with
  $O_1$ are used as a basis for the decomposition. Then, the coefficient of the right-nested commutator 
  $ [O_k, [O_{j},\ldots [O_i, O_1] \ldots ]]$ is the coefficient of the permutation $\alpha = (k j \ldots i 1)$
  in the original expression.
\end{itemize}  

These results can be readily applied to the expression (\ref{me.ite2}) for $\Omega_n$ by identifying $O_i = A(t_i)$. In
particular, for $\Omega_3$ a basis of right-nested commutators can be taken as $\{ [A(t_3),[A(t_2),A(t_1)]], [A(t_2),[A(t_3),A(t_1)]] \}$,
 associated with the permutations $(321)$ and $(231)$, respectively. The coefficients of $(3 2 1 )$ and $( 2 3 1 ) $ in (\ref{om.3})
are respectively $\frac{1}{3}$ and $ -\frac{1}{6}$, and so 
\[
  \Omega_3 = \frac{1}{3} \, A[3, 2, 1] - \frac{1}{6} \,  A[2, 3, 1] 
\]
in accordance with the previous direct calculation. 

Taking into account these considerations, we can write in general
\begin{equation}   \label{me.com1}
    \Omega_n(t) = \sum_{\sigma} \, (-1)^{d_b+1} \, \frac{d_a! (d_b + 1)!}{n!} \, 
    A[\sigma(2), \sigma(3), \ldots, \sigma(n),1],
\end{equation}
where now the sum extends over the $(n-1)!$ permutations $\sigma$ of $\{2, 3, \ldots, n \}$ and $d_a$ (respectively, $d_b$)
is the number of ascents (respect., descents) of the permutation $\sigma$ and thus $d_a + d_b = n-2$. Notice that the total number
of descents of the permutation $(\sigma(2) \, \sigma(3) \,\ldots \, \sigma(n) \,1)$ is precisely $d_b + 1$. Alternatively,
\begin{equation}   \label{me.com2}
\begin{aligned}
  & \Omega_n(t) = \frac{1}{n} \sum_{\sigma} \, (-1)^{d_b+1} \frac{1}{\binom{n-1}{d_b+1}} \, 
    \int_0^t dt_1 \int_0^{t_1} dt_2 \cdots \int_0^{t_{n-1}} dt_n \,  \\
    &  \qquad\qquad\qquad  [A(t_{\sigma(2)}), [A(t_{\sigma(3)}) \cdots 
    [A(t_{\sigma(n)}), A(t_1)] \cdots ]].
\end{aligned}    
\end{equation}
As an illustration, the expression of $\Omega_4$ reads
\begin{eqnarray*}  
  \Omega_4 & = & -\frac{1}{4} \, A[4, 3, 2, 1]  +  \frac{1}{12} \, A[4, 2, 3, 1] + \frac{1}{12} \, A[3, 2, 4, 1]  \\
 & & + \frac{1}{12} \, A[3, 4, 2, 1] +
  \frac{1}{12} \, A[2, 4, 3, 1] - \frac{1}{12} \, A[2, 3, 4, 1]. \nonumber
\end{eqnarray*}

According with the preceding results, one could
select any other $A(t_i)$ as the last operator (to the right) in the nested commutators, and so there are $n$ different but 
equivalent expressions for $\Omega_n$. In particular, if we take $O_1 \equiv A(t_n)$, then
\begin{equation}   \label{me.com3}
\begin{aligned}
  & \Omega_n(t) = \frac{1}{n} \sum_{\sigma \in \mathfrak{S}_{n-1}} \, (-1)^{d_b} \frac{1}{\binom{n-1}{d_b}} \, 
    \int_0^t dt_1 \int_0^{t_1} dt_2 \cdots \int_0^{t_{n-1}} dt_n \,  \\
    &  \qquad\qquad\qquad  [A(t_{\sigma(1)}), [A(t_{\sigma(2)}) \cdots 
    [A(t_{\sigma(n-1)}), A(t_n)] \cdots ]].
\end{aligned}    
\end{equation}
In any case, these identities 
 can be readily implemented in a computer algebra system to generate any order in the Magnus expansion.

\section{Concluding remarks}

The Magnus expansion is an extremely useful device when dealing with time-dependent linear differential equations of the form $Y' = A(t) Y$. It yields the solution of such equations
in exponential form, the exponent defined as an infinite series whose terms can be constructed in a recursive way as multiple integrals of nested commutators of the operator $A(t)$
defining the differential equation. Given its ubiquitous nature and the wide range
of applications in physics and mathematics, it is hardly surprising that  along the years several authors have proposed explicit formulas for the terms $\Omega_n(t)$
of the Magnus series. As
a matter of fact, the same formulas can be found in various published references, independently obtained by different authors. 
Such expressions could be classified into two types: either $\Omega_n$ is written as a
time-ordered integral of a sum of products of $A$ evaluated at different times (as in \cite{mielnik70cat}) or it is expressed as multiple integrals of a linear combination of $(n-1)$-nested commutators
\cite{iserles00lgm}. Of course, as pointed out in section \ref{sec.2}, by application of the Dynkin--Specht--Wever (DSW) theorem it is always possible to get an expression of the second type
from the first approach. The drawback, though, is that there are many redundancies due to the Jacobi identity and other identities of commutators appearing at high orders.

By contrast, in the procedure we propose here no use is done of the DSW theorem from (\ref{me.ite3}). Instead we apply the results obtained by Dragt \& Forest in \cite{dragt83con}
to get a general expression for $\Omega_n$ as an iterated integral of a linear combination of $(n-1)!$ right-nested independent commutators of $A$ evaluated at different times. 
Other expressions of this type have been obtained up to $\Omega_6$ containing less terms \cite{lakos17cef}, although no general expression has been proposed.

When developing our procedure we have also established
a remarkable connection of the Magnus expansion with the Malvenuto--Reutenauer Hopf algebra. This rather special Hopf algebra
 is non commutative, non cocommutative, free as an algebra, cofree as a 
coalgebra and self-dual \cite{hazewinkel10ara}. 
We have seen that the products defining this structure admits a natural interpretation in terms of 
products of the iterated integrals appearing in the
Magnus expansion, so this feature provides an additional, physical realization of the Hopf algebra of permutations. 

Given the close connection between the Magnus expansion and the Baker--Campbell--Hausdorff (BCH) formula (see e.g.
\cite{strichartz87tcb}), it is clear that the expression (\ref{me.com3}) can be used to get the homogeneous Lie polynomials
$Z_m(X,Y)$ in the expansion
\[
   Z = \log( \e^{X} \, \e^Y) = X + Y + \sum_{m=2}^{\infty} Z_m(X,Y).
\]
Proceeding in this way we recover the result obtained in \cite{loday94sdh}, although the resulting commutators appearing in
(\ref{me.com3}) are not all independent. An algorithm for expressing $Z$ in terms of a basis of the free algebra generated by
$X$ and $Y$ has been presented in \cite{casas09aea}.   

Although here we have treated only linear differential equations, it is clear that the same approach can also be applied to nonautonomous nonlinear systems with only
minimal changes \cite{strichartz87tcb,agrachev94vsa} and in fact to any problem where iterated integrals of the type considered in this work appear, such as the Wilcox expansion
in quantum mechanics \cite{wilcox67eoa},
chronological calculus in control theory \cite{agrachev79ter}, rough paths, etc.

Other issues remain of course to be analyzed in more detail, in particular the connection with other Hopf algebras closely related with the Malvenuto--Reutenauer Hopf algebra
such as the Hopf algebra of heap-ordered trees \cite{foissy13opf}, the role played by connected permutations \cite{poirier95adh} in our setting and the formulation at the level of the
Hopf algebra of the results obtained by Dragt \& Forest. This will be the subject of a forthcoming paper \cite{arnal17wip}.

After the completion of this work, we have become aware that the authors of  \cite{bandiera17eip} independently have obtained
expression
 (\ref{me.com3}) as a consequence of their treatment of the Euler idempotent based on the computation
 of a logarithm in a certain pre-Lie algebra of planar, binary, rooted trees.

\subsection*{Acknowledgments}
The authors are very grateful to Ander Murua for many insightful remarks, and especially for drawing their attention to the connection of the Magnus expansion with
the Hopf algebra of permutations. This work has been funded by the research project P1.1B20115-16 (Universitat Jaume I). FC has been additionally supported
by Ministerio de Econom\'{\i}a, Industria y Competitividad (Spain) through  project MTM2016-77660-P (AEI/FEDER, UE).

\bigskip

\bibliographystyle{plain}

\end{document}